\documentstyle{article}

\font\tenrm=cmr10
\font\tenit=cmti10
\font\elevenbf=cmbx10 scaled\magstep 1
\font\elevenrm=cmr10 scaled\magstep 1
\font\elevenit=cmti10 scaled\magstep 1

\def\ltap{\raisebox{-.4ex}{\rlap{$\sim$}} \raisebox{.4ex}{$<$}}

\renewenvironment{thebibliography}[1]
 { \elevenrm
   \begin{list}{\arabic{enumi}.}
    {\usecounter{enumi} \setlength{\parsep}{0pt}
     \setlength{\itemsep}{3pt} \settowidth{\labelwidth}{#1.}
     \sloppy
    }}{\end{list}}

\begin{document}
\vspace{-.3in}
\hfill\vbox{\hbox{\bf MAD/PH/758}
	    \hbox{May 1993}}\par
\vspace{-.2in}
\begin{center}
\vglue 0.6cm
{
 {\elevenbf        \vglue 10pt
               GENERAL ISSUES IN THE EVOLUTION\\
               \vglue 3pt
               OF FERMION MASSES AND MIXINGS
\\}
\vglue 1.0cm
{\tenrm V.~Barger, M.~S.~Berger\footnote{Talk presented by MSB at the
SUSY93 Conference, April 1, Northeastern University, Boston, MA.},
and P.~Ohmann \\}
\baselineskip=13pt
{\tenit Physics Department, University of Wisconsin\\}
\baselineskip=12pt
{\tenit Madison, WI 53706, USA\\}}

\vglue 0.8cm
{\tenrm ABSTRACT}

\end{center}

\vglue 0.3cm
{\rightskip=3pc
 \leftskip=3pc
 \tenrm\baselineskip=12pt
 \noindent
General issues in the renormalization group evolution of fermion masses and
mixings is discussed. An effective fixed point in the top quark Yukawa
coupling can strongly constrain its value at the electroweak scale. Predictions
following from Yukawa coupling unification are affected by threshold
corrections at the grand unified scale. The Landau pole translates into
an upper limit on the strong gauge coupling $\alpha _3(M_Z)$. Given the
hierarchy in the fermion sector, the evolution of the
Cabbibo-Kobayashi-Maskawa matrix can be expressed in terms of a single scaling
parameter $S$. Using this scaling factor and analogous scaling factors for
the quark and lepton masses, we outline a systematic strategy that readily
yields electroweak predictions for any GUT scale texture.}

\vglue 0.6cm
{\elevenbf\noindent 1. Introduction}
\vglue 0.2cm
\baselineskip=14pt
\elevenrm
The additional symmetry in grand unified theories (GUTs) can be used to
reduce the number of arbitrary parameters in the standard model. Gauge
coupling unification eliminates one of these free parameters. Yukawa coupling
unification$^1$ can potentially
provide a much more expansive reduction. In this
case, symmetries at the GUT scale can provide relations
between the 13 parameters of the flavor sector
(9 fermion masses and 4 parameters that characterize the mixing in the
Cabbibo-Kobayashi-Maskawa (CKM) matrix). From another point of view, the
low-energy measurements of fermion masses and mixings can provide a window
into the symmetries at the GUT scale.
In the following we will concentrate on a few general topics that are
relevant to the evolution of fermion masses and mixings.

\vglue 0.6cm
{\elevenbf\noindent 2. Fixed Points}
\vglue 0.2cm
Fixed point solutions$^{2-7}$
could apply for a wide range of top quark Yukawa couplings
arising in a more fundamental theory.
One can obtain a simple estimate of the location of the fixed point by
setting the one-loop top quark Yukawa
renormalization group equation$^8$ (RGE) in the minimal supersymmetric
standard model (MSSM) to zero,
\begin{eqnarray}
{{d\lambda _t}\over {dt}}&=&{{\lambda _t}\over {16\pi ^2}}
\Bigg (-\sum c_ig_i^2+6\lambda _t^2+\lambda _b^2
\Bigg )=0\;.
\end{eqnarray}
with $c_1=13/15$, $c_2=3$, $c_3=16/3$.
This is only accurate to about 10\% in practice
because the gauge couplings are themselves evolving.
A careful analysis of the two-loop RGEs in the MSSM using experimental
input for the gauge couplings yields an {\it effective} fixed point of
$\lambda _t^{fp}\simeq 1.1$ near the electroweak scale $\mu =M_Z$ as shown in
Figure 1. Top quark Yukawa couplings exceeding the fixed point value at the
GUT scale evolve rapidly to the fixed point, while the approach from below
is more gradual.

The prediction for the $m_b/m_{\tau }$
ratio provides motivation for the fixed point solution.
This behaviour can be
understood immediately from the one-loop RGE in the MSSM for
$R_{b/\tau}\equiv \lambda _b/\lambda _{\tau}$,
\begin{equation}
{{dR_{b/\tau}}\over {dt}}={{R_{b/\tau}}\over {16\pi ^2}}
\left (-\sum d_ig_i^2+\lambda _t^2
+3\lambda _b^2-3\lambda _{\tau}^2\right )\;,
\label{dRdt}
\end{equation}
with $d_1=-4/3$, $d_2=0$, $d_3=16/3$. If the
$b$-quark is sufficiently light and $R_{b/\tau }=1$ at the GUT scale,
large Yukawa couplings are required to
counteract the ``overshoot'' from the gauge coupling contributions
from Eq.~(2). Here we take
as inputs $m_{\tau }=1.784$ GeV and the running mass
$m_b(m_b)=4.25$ GeV$^9$.
In the standard model the effective fixed point solution
implies that the top quark is heavy $m_t > 200$ GeV. In the MSSM the
Yukawa coupling must be large ($\simeq 1$) at the electroweak scale, implying
a linear correlation between $m_t$ and $\sin \beta $ (neglecting contributions
from $\lambda _b$ and $\lambda _{\tau}$ which have a significant effect only
for very large $\tan \beta$),
\begin{eqnarray}
m_t(m_t)&=&{{\lambda _t^{fp}v\sin \beta }\over {\sqrt{2}}}=
{{\lambda _t^{fp}v}\over {\sqrt{2}}}{{\tan \beta}\over
{\sqrt {1+\tan ^2 \beta}}}\;,
\label{mt}
\end{eqnarray}
where $v=246$ GeV and $\tan \beta $ is the ratio of the vevs of the two
Higgs doublets in the
\vfill
{\baselineskip 12pt
\tenrm
\noindent Figure 1: The top quark Yukawa coupling evolves rapidly to the
effective fixed point value from above. The constraint $d\lambda _t/dt=0$
varies with scale because the gauge couplings are evolving.}
\newpage

\noindent MSSM.
As $\alpha _3(\mu )$ is increased, $\lambda _t(\mu )$ must be correspondingly
increased to preserve the $m_b/m_{\tau}$ prediction. Hence for larger
input $\alpha _3(M_Z)$, the solutions tend to display more strongly the fixed
point character.
The fixed point does {\it not} require that $\tan \beta $ be small,
but allows for large $\tan \beta $ if $m_t$ is sufficiently large.
There is  an intermediate region of
$\tan \beta $ in which the effects of $\lambda _b$ and $\lambda _{\tau }$
are negligible in the RG evolution, but for which Eq.~(\ref{mt}) is valid.
However if  $m_t^{\rm pole}$ is below $160$ GeV, the fixed point gives
$\tan \beta < 2$ with interesting consequences for Higgs boson
phenomenology$^{7}$.

Another interesting result is the observation$^3$ that the
observed $m_b/m_{\tau }$ ratio can be obtained if the masses of
all three members of the heavy generation are determined by fixed points
(without necessarily assuming the GUT scale unification constraint
$\lambda _b=\lambda _{\tau }$).
This solution requires that $\lambda _t$,
$\lambda _b$ and $\lambda _{\tau }$ be large, and
therefore that $\tan \beta $ be large. In some minimal models large
$\tan \beta $ will cause a violation of proton decay constraints.

\vglue 0.6cm
{\elevenbf\noindent 3. Threshold Corrections at the GUT Scale}
\vglue 0.2cm
Figure 2
shows the effects of taking threshold corrections to the GUT scale
unification constraint $\lambda _b(M_G^{})=\lambda _{\tau }(M_G^{})$ for two
different values of $\alpha _3(M_Z)$. The top quark mass plotted is the running
mass $m_t(m_t)$. For $\alpha _3(M_Z)=0.11$, the top
quark Yukawa coupling can be pushed below its fixed point for threshold
corrections as large as 20\%, and the solution of the RGEs is not
sufficiently close to the fixed point solution to provide a constraint
in the $m_t - \tan \beta $ plane. For $\alpha _3(M_Z)=0.12$ the fixed point
solution is useful even for large threshold corrections, since the solutions
display a stronger fixed point nature.

Threshold corrections to the GUT scale unification constraint
$\lambda _b=\lambda _{\tau }$ generally are larger if the
top quark Yukawa is large at the GUT scale;
however, Figure 2 indicates that these GUT threshold corrections
become less important in determining the relation between
$m_t$ and $\tan \beta$ for a large
top quark Yukawa coupling. It is precisely the fixed point nature of
$\lambda _t$ that make $m_t$,$\tan \beta$ solutions insensitive to even large
GUT threshold corrections. One also expects threshold
corrections to the other Yukawa coupling unification conditions, including
those involving the CKM mixing angles.

\vglue 0.6cm
{\elevenbf\noindent 4. Landau Pole}
\vglue 0.2cm
The two-loop part of the RGE's are known for the Yukawa couplings and for the
mixing angles in the MSSM. Comparing the two-loop to the one-loop can give
a quantitative estimate of the proximity of the Landau pole. Any criteria one
might define as the breakdown is admittedly subjective. We adopt one in
which the
two-loop contribution to the evolution be less than ${1\over 4}$ of the
one-loop contribution over the entire range of the Yukawa coupling evolution.
Since the top Yukawa coupling
is rising toward the Landau pole as one evolves
upward in scale, this condition is restrictive
at the highest scales.
The Landau pole indicates that there is an upper limit$^{4,6}$
on the value of the
strong coupling $\alpha _3(M_Z)\ltap 0.125$.

\newpage
\vspace*{19cm}
{\baselineskip 12pt
\tenrm
\noindent Figure 2: The effect of threshold corrections on the Yukawa coupling
unification condition $\lambda _b(M_G^{})=\lambda _{\tau }(M_G^{})$
with $m_b=4.25$ GeV for
$\alpha _3(M_Z)=0.11$ and $0.12$. The corrections have a more pronounced effect
for smaller values of $\alpha _3(M_Z)$ for which the solutions are closer to
the fixed point. The Landau pole provides a constraint on corrections with
$\lambda _b(M_G^{})>\lambda _{\tau }(M_G^{})$.}

\newpage
Figure 1 also shows that for $\lambda _b(M_G^{}) > \lambda _{\tau}(M_G^{})$,
the top quark
Yukawa coupling is pushed up against the Landau pole. This potentially
can give new constraints on the size of GUT scale threshold corrections.

\vglue 0.6cm
{\elevenbf\noindent 5. Universal Evolution of the CKM Matrix}
\vglue 0.2cm
The one-loop evolution equations for the Yukawa coupling matrices
are$^{10-15}$
\begin{eqnarray}
{{d{\bf U}}\over {dt}}&=&{1\over {16\pi ^2}}
\Big (x_u{\bf I}+y_u{\bf UU^{\dagger }}+a_u{\bf DD^{\dagger }}
\Big ){\bf U} \;,
\label{dUdt}
\end{eqnarray}
\begin{eqnarray}
{{d{\bf D}}\over {dt}}&=&{1\over {16\pi ^2}}
\Big (x_d{\bf I}+y_d{\bf DD^{\dagger }}+a_d{\bf UU^{\dagger }}
\Big ){\bf D} \;,
\label{dDdt}
\end{eqnarray}
where the coefficients $x_i$, $y_i$, $a_i$
depend upon the particle content of
the theory and are functions of the dimensionless
gauge and Yukawa couplings, i.e.
$a_i=a_i(g_1^2,g_2^2,g_3^2,
{\bf Tr}[{\bf UU^{\dagger }}],{\bf Tr}[{\bf DD^{\dagger }}],
{\bf Tr}[{\bf EE^{\dagger }}])$ and Higgs quartic couplings.
When there is a hierarchy of masses in the Yukawa matrices, the evolution of
the quark masses and CKM mixing angles is given as a simple scaling.
The hierarchy required is the following: Light generations with small
Yukawa couplings (i.e. $<<1$), and a heavy third generation.
With such a hierarchy it will be these heavy Yukawas along with the gauge
couplings that are important in the CKM evolution.
Mixing between the heavy and
light generations must be small, which occurs naturally for a hierarchy
in the Yukawa matrices.
The only terms in Eqs.~(4) and (5) that contribute to the running of the
CKM matrix are the ones involving $a_u$ and $a_d$. The CKM
mixing angles scale as$^{13-15}$
\begin{equation}
{{dW_1}\over {dt}}=-{{W_1}\over {8\pi ^2}}
\left (a_d\lambda _t^2
+a_u\lambda _b^2\right ) \;, \label{dW1dt}
\end{equation}
where $W_1=|V_{cb}|^2, |V_{ub}|^2, |V_{ts}|^2, |V_{td}|^2$, the
CP-violation parameter$^{16}$ $J$
and
\begin{equation}
{{dW_2}\over {dt}}=0\;, \label{dW2dt}
\end{equation}
where $W_2=|V_{us}|^2, |V_{cd}|^2, |V_{tb}|^2, |V_{cs}|^2, |V_{ud}|^2$.
The two-loop versions of Eqs.~(4)-(7) can be found in Ref.~\cite{BBO2}.
The solution of Eq.~(6) is
\begin{equation}
W_1(M_G^{})=W_1(\mu )S(\mu )\;,
\end{equation}
where $S$ is a scaling factor defined by
\begin{equation}
S(\mu ) = \exp \left \{ -{1\over {8\pi ^2}}\int\limits_{\mu}^{M_G}
\left (a_d\lambda _t^2
+a_u\lambda _b^2\right )\;
d\ln\mu' \right \}
\label{defS} \;.
\end{equation}
The lightest two generations do not affect the evolution, and
one does not need the mixing between the first two generations to be small
for the universal scaling described above to occur.
This makes the scaling universality an especially good approximation
since the Cabbibo angle is the largest of the quark mixings.
Any amount of mixing between light generations is allowed,
which is intuitively the
case since they have a negligible impact on the evolution.
The scaling behavior can be demonstrated to all orders in perturbation
theory$^{14}$.

Corollaries to the universality of the scaling of the CKM matrix are
the following:

\begin{itemize}
\item The ratios $\left |V_{ub}/V_{cb}\right |$,
$\left |V_{td}/V_{ts}\right |$
do not scale.

\item The CP-violation parameter $J$ has the same scaling factor $S$.
There is a simple way to understand the scaling of $J$ in terms of the
mixing angles. The unitarity relation
\begin{eqnarray}
V_{11}^{}V_{12}^*+V_{21}^{}V_{22}^*+V_{31}^{}V_{32}^*&=&0\;, \label{unit}
\end{eqnarray}
can be represented by the triangle in Figure 3a. Since
$|V_{11}^{}V_{12}^*|\approx |V_{21}^{}V_{22}^*|>>|V_{31}^{}V_{32}^*|$, this
particular unitarity triangle is very slim. Only the short side
scales to leading order in the approximation, as shown in Figure 3b.
Since the CP-violation parameter $J$ is twice the area of any unitarity
triangle, it must scale with the same factor $S$. Of course the other sides
of the triangle must change a very small amount to preserve unitarity
\begin{eqnarray}
V_{11}^{}V_{12}^*+V_{21}^{}V_{22}^*+SV_{31}^{}V_{32}^*&=&0\;, \label{units}
\end{eqnarray}
but these changes are subleading in the hierarchy approximation.
A similar argument exists
for the other unitarity triangles.
Equivalent ways to think about the scaling is in terms of the Wolfenstein
parameterization$^{18}$ or the
DHR parameterization$^{17,19}$.
In the former case the scaling manifests itself as the running of only $A$, and
the nonevolution of $\lambda $, $\rho $, and $\eta $. In the latter case,
the mixing angles $s_1$ and $s_2$ do not scale, while $s_3$ scales with the
factor $S^{1/2}$.

\item If the mixing between two generations is exactly zero, then it must be
zero at all scales. This is true even if there is no hierarchy.

\end{itemize}
\vfill
\begin{center}
{\baselineskip 12pt
\tenrm
\noindent Figure 3: Scaling of a unitarity triangle.}
\end{center}

\newpage

A general texture analysis can be performed by diagonalizing mass
matrices at the scale where they are simple (i.e. at the GUT scale
where the zero structure is defined).
The largest corrections from subleading terms in the hierarchy will come at
this stage (they can be as large as ${\cal O}(\lambda ^2)\simeq 5\%$).
The contribution of the subleading terms
to the RGEs that are neglected in the hierarchy approximation is
much smaller and can be neglected entirely.

A crucial point to be emphasized here is that the mass matrices
themselves
contain more information than can be observed. To compare predictions with
experiment it is only necessary to evolve the observables, i.e. the masses
and mixings. The zeroes disappear from the mass matrices as the
low energy theory does not respect the discrete (or otherwise) symmetries
that gave rise to them, but the evolution of the observables is particularly
simple given that the hierarchy exists.

A practical, systematic strategy to generate the electroweak predictions
of various GUT textures is the following:

(1) There are scaling quantities for the heavy and light Yukawa couplings and
for the CKM matrix that depend on the values of the heavy Yukawa couplings and
the gauge couplings.
For any particular choice of these couplings at the electroweak scale there
are particular solutions for the scaling parameters that can be calculated
using the RGEs. For example, after fixing the gauge couplings at the
electroweak scale, contours
of the scaling factors can be obtained in the $m_t$,$\tan \beta $ plane.

(2) For any given texture find the diagonal Yukawa couplings,
the CKM matrix, and the
parameter $J$ at the ``texture'' or GUT scale.  One can retain
the contributions of the subleading terms in the diagonalization
to any degree of accuracy. These contributions (if desired)
can be obtained analytically(e.g. Ref.~\cite{Giudice})
or numerically if necessary.

(3) The evolution of the observables calculated in step (2) can now be evolved
to the electroweak scale
by multiplying by the scaling factors calculated in step (1).

(4) Step (2) can be repeated for a different texture to obtain a
different set of boundary conditions. The scaling factors from step (1)
are obtained from the evolution equations alone and {\it need not} be
recalculated. The evolution for the new texture is obtained by
simply multiplying
by the scaling factors in step (1).

A more sophisticated algorithm is needed to evolve the Yukawa
matrices as a whole. This extra work is unnecessary, however. Only the
observable components need to be evolved, and this can be done without
even making the hierarchy approximation. However, then the evolution
is only approximately described by scaling
(although the approximation is quite good).
The complete two-loop evolution
equations for the mixing angles and the quark masses are known for any
theory in which the RGEs for the Yukawa matrices are known$^{14}$.
Each entry in the Yukawa matrices is not known after evolution, but the
observable combinations are known and this is the full information needed to
compare with experiment.

\vglue 0.6cm
{\elevenbf\noindent 6. Conclusions}
\vglue 0.2cm
Low energy observables in the flavor sector can provide a powerful probe
of the GUT scale symmetries. The most predictive models provide an enormous
reduction in the number of arbitrary parameters. Even if these models are
eventually in contradiction with improvements in experimental data, we are
confident that the renormalization group scaling of low energy
observables will continue to be a valuable tool in the search for higher
symmetries.

\vglue 0.6cm
{\elevenbf\noindent 7. Acknowledgements}
\vglue 0.2cm
We wish to thank M.~Peskin for emphasizing the importance of threshold
corrections in Yukawa unification. This research was supported
in part by the University of Wisconsin Research Committee with funds granted by
the Wisconsin Alumni Research Foundation, in part by the U.S.~Department of
Energy under contract no.~DE-AC02-76ER00881, and in part by the Texas National
Laboratory Research Commission under grant no.~RGFY9273.
PO was supported in part by an NSF Graduate Fellowship.

\vglue 0.6cm
{\elevenbf\noindent 8. References}
\vglue 0.2cm

\end{document}